\documentclass[conference]{IEEEtran}
\IEEEoverridecommandlockouts
\usepackage{cite}
\usepackage{listings}
\usepackage[colorinlistoftodos]{todonotes}
\usepackage{amsmath,amssymb,amsfonts}
\usepackage{algorithmic}
\usepackage{graphicx}
\usepackage{textcomp}
\usepackage{xcolor}
\usepackage[T1]{fontenc}

\def\BibTeX{{\rm B\kern-.05em{\sc i\kern-.025em b}\kern-.08em
    T\kern-.1667em\lower.7ex\hbox{E}\kern-.125emX}}
\begin{document}

\title{GPT-2C: A GPT-2 parser for Cowrie honeypot logs\\
{\footnotesize \textsuperscript{}}
}

\author{\IEEEauthorblockN{1\textsuperscript{st} Febrian Setianto}
\IEEEauthorblockA{\textit{Center for Ubiquitous Computing} \\
\textit{University of Oulu}\\
Oulu, Finland \\
febrian.setianto@oulu.fi}

\and
\IEEEauthorblockN{2\textsuperscript{nd} Erion Tsani}
\IEEEauthorblockA{\textit{Computer Engineering \& Informatics} \\
\textit{University of Patras}\\
Patra, Grece \\
eriontsani01@gmail.com}

\and
\IEEEauthorblockN{3\textsuperscript{rd} Fatima Sadiq}
\IEEEauthorblockA{\textit{Center for Ubiquitous Computing} \\
\textit{University of Oulu}\\
Oulu, Finland \\
fatima.sadiq@oulu.fi}

\and
\IEEEauthorblockN{4\textsuperscript{th} Georgios Domalis}
\IEEEauthorblockA{\textit{Computer Engineering \& Informatics} \\
\textit{University of Patras}\\
Patra, Grece \\
domalis@ceid.upatras.gr}

\and
\IEEEauthorblockN{5\textsuperscript{th} Dimitris Tsakalidis}
\IEEEauthorblockA{\textit{NOVELCORE} \\
Patra, Grece \\
tsakalidis@novelcore.eu}

\and
\IEEEauthorblockN{6\textsuperscript{th} Panos Kostakos}
\IEEEauthorblockA{\textit{Center for Ubiquitous Computing} \\
\textit{University of Oulu}\\
Oulu, Finland \\
panos.kostakos@oulu.fi}
}

\maketitle
\begin{abstract}
Deception technologies like honeypots produce comprehensive log reports, but often lack interoperability with EDR and SIEM technologies. A key bottleneck is that existing information transformation plugins perform well on static logs (e.g. geolocation), but face limitations when it comes to parsing dynamic log topics (e.g. user-generated content). In this paper, we present a run-time system (GPT-2C) that leverages large pre-trained models (GPT-2) to parse dynamic logs generate by a Cowrie SSH honeypot. Our fine-tuned model achieves 89\% inference accuracy in the new domain and demonstrates acceptable execution latency.

\end{abstract}


\section{Introduction}
The continuously evolving landscape of cyberspace attacks has led to advanced defence strategies in the area of network forensics. These include the use of automated agents with the ability to make knowledgeable decisions and plan dynamic responses, as well as the use of deception techniques and tools, including honeypots and honeynets, which can detect, prevent and identify malicious activities.

Honeypots have evolved from simple low-interaction systems simulating specific parts of real systems to high-interaction systems that are set up to appear as complete operating systems with real network services, real or virtual devices and realistic information of users \cite{8635603}. Their main role is to engage attackers and captivate their behavior and practices so that operators can analyze in-depth the vulnerabilities of their systems \cite{1254322}. Honeynets are networks of interconnected honeypots operating as a whole alongside a functional Intrusion Detection System (IDS) \cite{franco2021survey}. 

Cowrie is a virtual Secure Shell (SSH) and Telnet honeypot, utilized to continuously monitor brute force attacks, by producing log files capturing the actions of malicious users. However, they can be easily identified by adversaries, since they are rarely customized to systems, due to the lack of standard frameworks \cite{9071014}. To better understand and leverage the deceptive capabilities of Cowrie, it is important to automate the parsing methodologies of the log files produced.

\section{Towards AI-based parsing}

Unstructured information describes data as a large text-corpus collection with no predefined model. Massive volumes of unstructured information is extracted from various domains in a variance of forms, from media, texts to log files. They can be processed and parsed for semantic search \cite{kostakos2020strings}, anomaly detection \cite{7774521}, cognitive management \cite{8255757}, autonomous problem validation \cite{BeschastnikhBSSE2011}, incident management or root cause analysis \cite{6606586}. Information extraction from unstructured data is enabled through (i) relation-based extraction, (ii) template-based extraction, (iii) deep learning-based extraction\cite{6606586,10.1145/500737.500757}.

Relationship-based extraction requires linguistic data. Named entity recognition and text classification along with Conditional Random Fields (CRFs) enable the identification of patterns in word sequencing leading to the recognition of entities by the pattern of sequence containing it \cite{SuhLee2016MiningUL}. Another relationship-based approach focuses on the metadata extraction by utilizing a part of speech tagging methodology followed by a morphological analysis that identifies nested words and assigns them labels based on its context towards extracting features \cite{10.1145/500737.500757}.

In contrast to relationship-based extraction, template-based information extraction utilises various data mining techniques and is used mostly on non-linguistic unstructured data. Agent-based approaches that analyse data based on a unique protocol classifies it and extracts table templates in heterogeneous sources, but are effective on specific message corpora \cite{5360240}. Pattern learning techniques have been widely adopted for template-based information extraction. A widely used open-source algorithm, namely the Simple Logfile Clustering Tool (SLCT), works in three steps: word vocabulary construction and capture word frequency, clustering the inputs from word vocabulary, and construction of the template through the log messages\cite{1251233}.

Deep learning techniques used for data analysis and information extraction are solely dependent on data. DeepDeSRT, a deep learning neural network performs image analysis to identify table formats, constructs templates \cite{8270123}. On the other hand, DeepLog enables Long Short-Term Memory (LSTM) to recognize word distribution patterns rather than analyse data and extract information\cite{10.1145/3133956.3134015}.

In general, parsers struggle with the domain/vendor-specific technical language, while developing and maintaining them is a time-consuming process with high demanding domain expertise. An autonomous log parser to be (i) resilient to different data formats, (ii) eliminate the need for human intervention, (iii) independent from labelled training data, (iv) does not require a dictionary to parse different table components such as rows, columns, and headers\cite{9370654}.

Generative Pre-trained Transformer (GPT) models by OpenAI \cite{OpenAI} have taken the natural language processing (NLP) community by storm by introducing very powerful language models. These models can perform various NLP tasks like question answering, textual entailment, and text summarisation without any supervised training.

GPT-2  is a transformers pre-trained model \cite{wolf-etal-2020-transformers} on a very large corpus of internet text data in English, more precisely a dataset of 8 million web pages (WebText). The model was trained to \emph{guess the next word in sentences}, so it performs best on the pre-trained domain, however, it performs well on downstream tasks. In this paper, we present a GPT-2 based approach that leverages deep learning to automate the process of extracting and parsing information from unstructured log files to better understand and leverage the deceptive capabilities of Cowrie.

\section{Methods}
GPT-2C is a run-time tool that leverages a fine-tuned GPT-2 model to help overcome the bottleneck of dynamic log reports generated by Cowrie. Trained on cowrie log data, the model enables us to essentially translate a typical data transformation task into a Question Answering (QA) problem. 

\subsection{Training Data}
We fine-tuned a GPT-2 model on the CyberLab honeynet dataset which is freely available in Zenodo \cite{sedlar_urban_2020_3687527}. The dataset contains Cowrie honeypot logs with attributes in JSON format from May 2019 to February 2020. For the purposes of our experiment, we extracted only the \verb#message# attribute which is the most dynamic topic in the log output, containing Linux commands and utilities used during attacks. To minimise computing time, we collected data from 29 randomly selected JSON log files out of the total 290. Next, we explain how we fine-tuned our model following the required QA format using the Hugging Face library\cite{wolf-etal-2020-transformers}.

\subsection{Fine-tuning GPT-2 log parser: Q\&A model}

For the experiment, we selected the GPT-2 small model with 117M parameters, 12 layers and 1024 dimensionality as GPT-2 can only generate a maximum of 1024 tokens per request. Furthermore, the GPT-2 small size model is easy to fine-tune and the weight of the model can adapt better to honeypot logs data\cite{radford2019language}. Our data contains 36,445 unique commands and was split into training (32,801), validation (3,645) and test (999) sets. We used two epochs and a batch size of two, depending on the maximum sequence length and GPU memory. 

\subsection{Evaluation Metrics}
For evaluation metrics, in addition to the loss function that the base GPT-2 provides, we also calculated the F1-score, a common metric for NLP tasks. The formula for the standard F1-score is the harmonic mean of the precision and recall. A perfect model has an F-score of 1.

$$
F_1 = 2*\frac{precision * recall}{precision + recall} = \frac{tp}{tp + 1/2(fp+fn)} 
$$

{\small\emph{where tp stands for true positive, fp for false positive and fn for false negative} } \\
 
In the Span-based QA task we compared the tokens of the correct answer and the predicted. To achieve this, we have reused the methods developed for evaluating models targeting the Stanford Question Answering Dataset (SQuAD)\cite{medium.com}. \\

\subsection{The Cowrie SSH Honeypot}

The Cowrie instance was set up on top of Ubuntu 20.04.2 LTS and deployed in the cloud. We changed the actual SSH port to establish a remote connection since the default port (22) is employed by Cowrie. Next, we set the firewall to permit incoming connections towards both ports. Additionally, we replaced several default values from Cowrie's configuration to diminish the likelihood of attackers acknowledging the honeypot environment. Some of the variations including hostname, Linux distribution, user accounts, CPU information and mounted media. While Cowrie produced log files in plain text and JSON formats, the subsequent processing pipeline only took the structured JSON logs.

\subsection{Transformer Pipeline architecture}

\begin{figure}[h!]
\centerline{\includegraphics[width=0.3\textwidth]{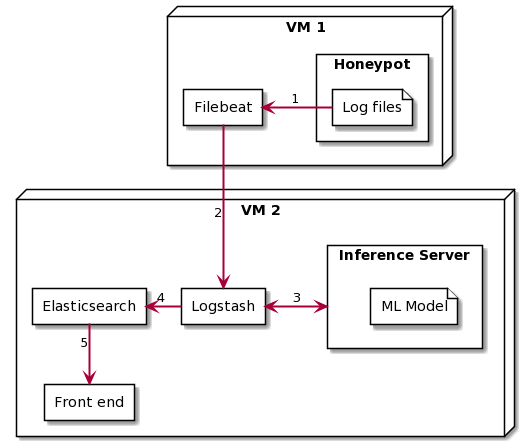}}
\caption{End-to-end pipeline architecture, with arrows depicting data flow.}
\label{architecture}
\end{figure}

We developed a pipeline to deploy the fine-tuned GPT-2 model and tested it on real-time logs produced by the Cowrie honeypot.
Our testing environment was set up on top of Ubuntu 20.04.2 LTS and comprised of two nodes as illustrated in Fig.~\ref{architecture}.
The first node, VM 1, is deployed on an instance with 4GB of RAM and 3 CPU cores, as the purpose is to serve as a honeypot.
The second node is equipped with 112GB of RAM, 14 CPU cores, and an NVIDIA Tesla P100.
All the nodes were deployed in cloud  infrastructure which is based on OpenStack.

Several components are part of the building blocks of this system.
First is Cowrie, the honeypot instance which records the intruders' actions, timestamp, and several other fields.
The second is Filebeat which is responsible for collecting Cowrie's log files and harvesting the content.
The third is Logstash which process each log message or line as a JSON object. It also parsed the timestamp field with ISO8601 format. It then removed unnecessary information, such as log file location, and send the attackers' commands to an Inference Server if the message satisfies a condition.
The fourth is the Inference Server, which is constructed on top of FastAPI. It provides an endpoint that accepts the attacker's commands, feeds them to the machine learning model to perform the inference, and returns the prediction result with the score as the response.
The fifth is Elasticsearch which stores and indexes the log messages given by Logstash to support data exploration and aggregation.
The last is the front-end that provides an easy to use user interface for data exploration and visualization, powered by Kibana.

\subsection{Run-time transformer parser: FastAPI to deploy the GPT-2 model}

Instead of mandating Logstash to load the ML model and perform the inference on each of the attackers' commands, the responsibility is lifted to a dedicated entity named Inference Server.
This design allows separation of concern: Logstash utilized its resources to parse and transform the logs, and Inference Server to load the ML model and serve incoming requests concurrently.

The Inference Server was built on top of FastAPI\cite{fastapi}, a fast web framework for building web APIs.
A Python-based framework was chosen to support the inference, as the ML model itself was trained and saved according to PyTorch's format\cite{pytorch}.
It exposed a RESTful API over HTTP protocol, where it only accepts a request with JSON-formatted payload.

\begin{figure}[h!]
\centerline{\includegraphics[width=0.3\textwidth]{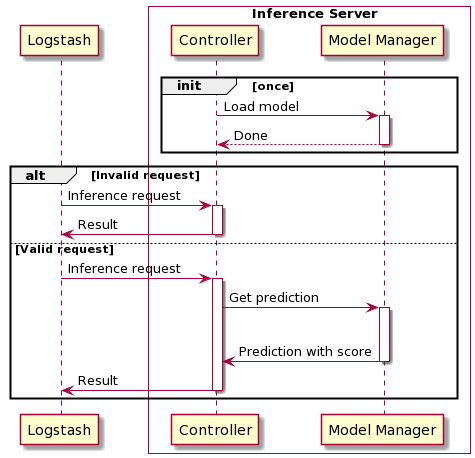}}
\caption{The interaction among Logstash, Inference Server, and ML model.}
\label{inference-server}
\end{figure}

When Logstash makes the HTTP request, Controller first checks the request's validity.
An invalid request then will be rejected with an appropriate response and status code.
On the contrary, Controller forwards a valid request payload to the Model Manager.
The Model Manager will then accept the string, invoke the inference function, and return the prediction result along with the confidence score.
Next, the Controller will wrap the results in a Python \verb#dictionary#, and it will convert it into a JSON-formatted response.
The Model Manager only loaded the ML model once when the Server was starting up.
This method allows the inference process to use the once-loaded model from memory.
Fig.~\ref{inference-server} illustrates the complete process.

\section{Results}
\begin{figure}[h!]
\centerline{\includegraphics[width=0.4\textwidth]{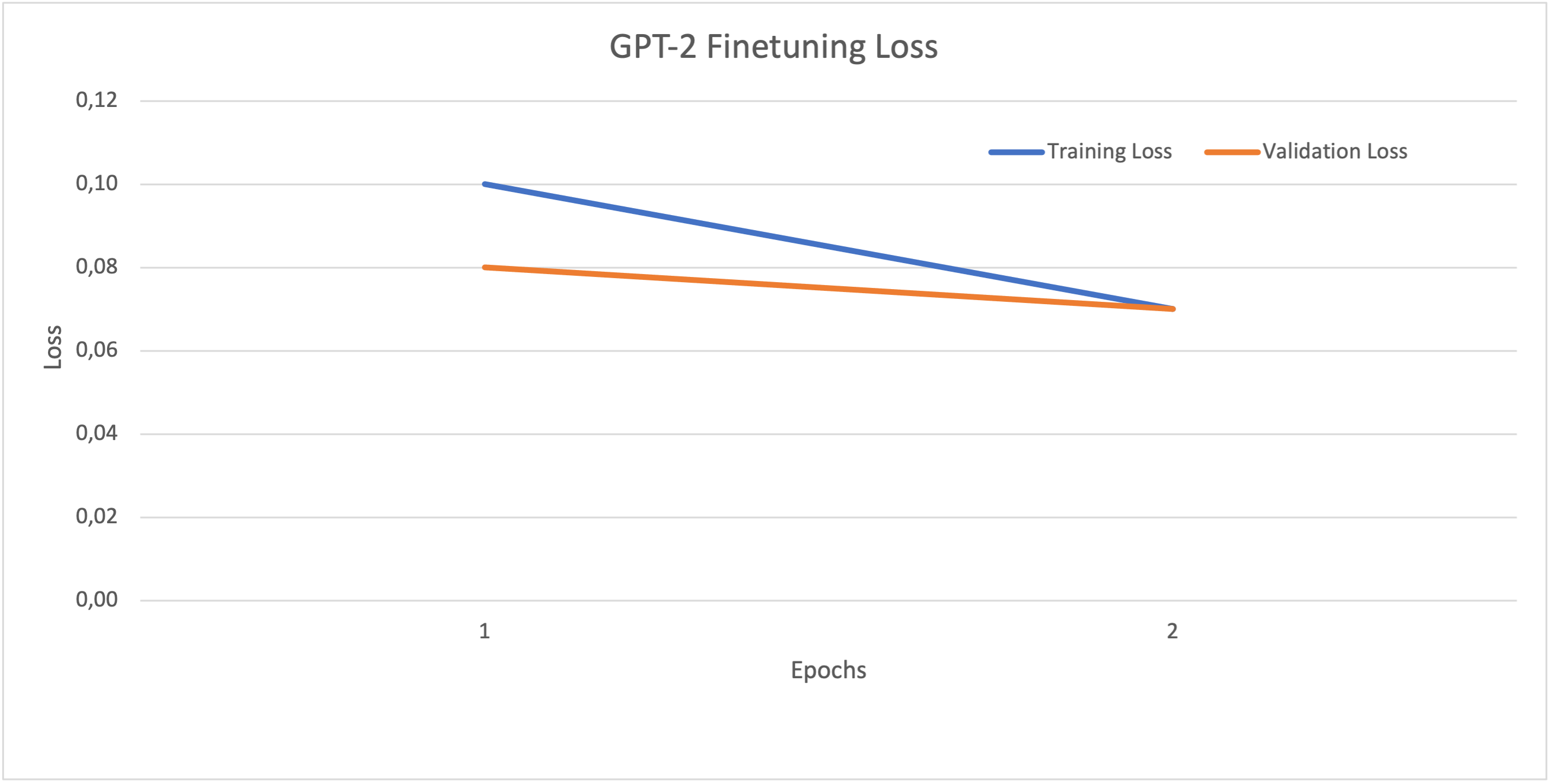}}
\caption{Training loss and validation loss curves.}
\label{lossgpc}
\end{figure}

Following prior work \cite{medium.com}, in order to gauge the performance of our system, our evaluation task is a span-based Question Answering task, where the context is an array of utilities and the answer is the utility predicted by our model. As shown in Table~\ref{tab1}, the proposed approach returns an accuracy of 89\% (F1-score) on the out-of-sample test set. The dual learning curves in Fig.\ref{lossgpc} suggest a good fit, with the training and validation loss ranging from 0.07 to 0.10.

\begin{table}[htbp]
\caption{GPT-2 Model Generated Results}
\begin{center}
\begin{tabular}{|c|c|}
\hline
\textbf{Epochs/Batch size} & 2   \\
\hline
\textbf{Training Samples} & 32801   \\
\hline
\textbf{Longest Token} & 783    \\
\hline
\textbf{Validation samples} & 3645  \\
\hline
\textbf{Overall time taken} & 2:54  \\ 
\hline
\textbf{testing f1 score} & 0,89827  \\ 
\hline
\end{tabular}
\label{tab1}
\end{center}
\end{table}

\begin{figure}[h!]
\centerline{\includegraphics[width=0.5\textwidth]{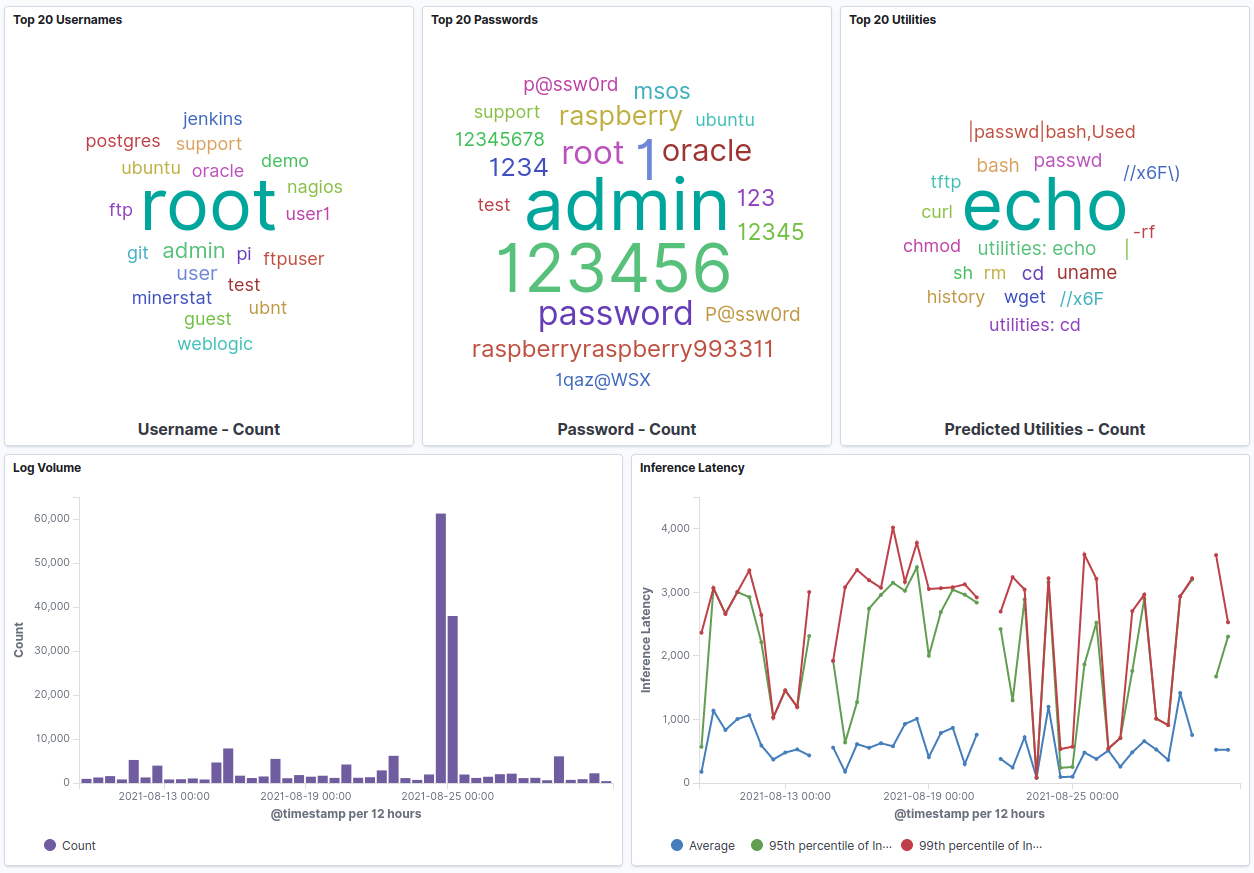}}
\caption{Kibana as the front-end supports various visualization types.}
\label{dashboard}
\end{figure}

To evaluate the performance of the model at run-time, we configured Elasticsearch to collect various metrics from our Cowrie instance. Performance metrics are subsequently visualised in a Kibana dashboard shown in Fig.~\ref{dashboard}. The explanation of each panel from the top left to bottom right is as follows:

\begin{enumerate}
    \item \textbf{Top left:} Top 20 usernames that the attackers have used.
    \item \textbf{Top middle:} Top 20 passwords that the attackers have used. 
    \item \textbf{Top right:} Top 20 tools predicted from ML model. Inference result from leveraging the ML model.
    \item \textbf{Bottom left:} Log volume per 12 hours bucket. Counted from each log stored in Elasticsearch.
    \item \textbf{Bottom right:} Average, 95th and 99th percentiles of inference latency. 1 in 100 inference process could take up to 4 seconds. Blanks between the lines indicate the temporary lack of attackers' activities.
\end{enumerate}

\section{Conclusion}
In this paper, we present a prototype that leverages large pre-trained models for parsing dynamic log-lines from a Cowrie SSH honeypot. Our experiments have focused on parsing utilities and commands that are often abused by malware, and have achieved good accuracy on out-of-sample validation set. Furthermore,  system metrics indicate an acceptable inference latency on run-time. The simplicity of the proposed system and the rapid conversion of the GPT-2 model on the specific domain can facilitate the adaptation of AI-driven solutions in EDR and SIEM technologies. 

\section*{Acknowledgment}
This research work has been financially supported by EU Horizon 2020 project GLASS (959879), EU Horizon 2020 project IDUNN (101021911) and by Academy of Finland 6Genesis Flagship (318927).

\bibliographystyle{plain}
\bibliography{bibliography.bib}

\end{document}